\documentclass[aps, prl, amsmath, floats,floatfix, twocolumn, superscriptaddress, nofootinbib, showpacs]{revtex4}

\usepackage{amssymb}
\usepackage{amsmath}
\usepackage{verbatim}
\usepackage{mathrsfs}
\usepackage{amsfonts}
\usepackage{latexsym}
\usepackage{epsfig}
\usepackage{color}
\usepackage{graphicx,subfigure}
\usepackage{units}
\usepackage{slashbox}

\begin{document}


\definecolor{orange}{rgb}{0.9,0.45,0}

\newcommand{\re}{\mbox{Re}}
\newcommand{\im}{\mbox{Im}}

\newcommand{\argelia}[1]{\textcolor{pink}{{\bf Argelia: #1}}}
\newcommand{\dario}[1]{\textcolor{red}{{\bf Dario: #1}}}
\newcommand{\juanc}[1]{\textcolor{green}{{\bf JC: #1}}}
\newcommand{\juan}[1]{\textcolor{cyan}{{\bf Juan B: #1}}}
\newcommand{\alberto}[1]{\textcolor{blue}{{\bf Alberto: #1}}}
\newcommand{\miguela}[1]{\textcolor{red}{{\bf Miguel: #1}}}
\newcommand{\miguelm}[1]{\textcolor{orange}{{\bf Megevand: #1}}}
\newcommand{\OS}[1]{\textcolor{magenta}{{\bf Olivier: #1}}}

\renewcommand{\t}{\times}

\long\def\symbolfootnote[#1]#2{\begingroup%
\def\thefootnote{\fnsymbol{footnote}}\footnote[#1]{#2}\endgroup}


\title{Schwarzschild black holes can wear scalar wigs}

\author{Juan Barranco} 
\affiliation{Departamento de F\'isica, Divisi\'on de Ciencias e Ingenier\'ia, Campus Le\'on, 
Universidad de Guanajuato, Le\'on 37150, M\'exico}

\author{Argelia Bernal} 
\affiliation{Instituto de Ciencias Nucleares, Universidad Nacional
  Aut\'onoma de M\'exico, Circuito Exterior C.U., A.P. 70-543,
  M\'exico D.F. 04510, M\'exico}

\author{Juan Carlos Degollado} 
\affiliation{Instituto de Astronom\'{\i}a, Universidad Nacional
  Aut\'onoma de M\'exico, Circuito Exterior C.U., A.P. 70-264,
  M\'exico D.F. 04510, M\'exico}

\author{Alberto Diez-Tejedor}
\affiliation{Departamento de F\'isica, Divisi\'on de Ciencias e Ingenier\'ia, Campus Le\'on, 
Universidad de Guanajuato, Le\'on 37150, M\'exico}

\author{Miguel Megevand}
\affiliation{Instituto de Ciencias Nucleares, Universidad Nacional
  Aut\'onoma de M\'exico, Circuito Exterior C.U., A.P. 70-543,
  M\'exico D.F. 04510, M\'exico}

\author{Miguel Alcubierre}
\affiliation{Instituto de Ciencias Nucleares, Universidad Nacional
  Aut\'onoma de M\'exico, Circuito Exterior C.U., A.P. 70-543,
  M\'exico D.F. 04510, M\'exico}

\author{Dar\'{\i}o N\'u\~nez}
\affiliation{Instituto de Ciencias Nucleares, Universidad Nacional
  Aut\'onoma de M\'exico, Circuito Exterior C.U., A.P. 70-543,
  M\'exico D.F. 04510, M\'exico}

\author{Olivier Sarbach}
\affiliation{Instituto de F\'{\i}sica y Matem\'aticas, Universidad
Michoacana de San Nicol\'as de Hidalgo, Edificio C-3, Ciudad
Universitaria, 58040 Morelia, Michoac\'an, M\'exico}


\date{\today}


\begin{abstract} 
We study the evolution of a massive scalar field surrounding a
Schwarzschild black hole and find configurations that can survive for
arbitrarily long times, provided the black hole or the scalar field
mass is small enough.  In particular, both ultra-light scalar field
dark matter around supermassive black holes and axion-like scalar
fields around primordial black holes can survive for cosmological
times.  Moreover, these results are quite generic, in the sense that
fairly arbitrary initial data evolves, at late times, as a combination
of those long-lived configurations.
\end{abstract}




\maketitle


{\it I- Introduction.}  It has been known for some time that a
Schwarzschild black hole (BH) cannot support a nontrivial scalar field
(SF) distribution, i.e. such a hole does not admit scalar
hair~\cite{Bekenstein:1995un, Pena:1997cy}.  However, the no-hair
theorems do not exclude the existence of dynamical solutions that
decay very slowly in time.  In a recent work~\cite{Burt:2011pv} we
found regular SF configurations surrounding a Schwarzschild BH that
can survive for relatively long times, and conjectured that, for a
certain range of values of the SF and BH masses, such SF
configurations could survive in the vicinity of the BH for
cosmological timescales.  Here we show that this is indeed what
happens, and that such distributions arise from the evolution of
fairly arbitrary initial configurations.  Our results give support to
the idea that the dark matter halos could be described by a coherent
scalar excitation~\cite{Sin:1992bg, Arbey:2001qi}
(see~\cite{UrenaLopez:2002du, CruzOsorio:2010qs} for papers with a
similar motivation), and to the possibility of nontrivial axion
distributions surrounding primordial BHs~\cite{Sikivie:2009qn}.


{\it II- Three types of resonances.}  The dynamics of a SF propagating
on a Schwarzschild background are described by the Klein-Gordon
equation
\begin{subequations}\label{KG.Pot}
\begin{equation}
\frac{\partial^2\phi}{\partial t^2} - \frac{\partial^2\phi}{\partial r_*^2}
+ V(r)\phi = 0 \; ,
\label{Eq:KG}
\end{equation}
with $V(r)$ an effective potential given by
\begin{equation}
V(r) = \left(1 - \frac{2M}{r} \right)\left( \frac{\ell(\ell+1)}{r^2}
 + \frac{2M}{r^3} + \mu^2 \right) \; .
\label{Eq:KGPotential}
\end{equation}
\end{subequations}
Here $M$ is the BH mass, $\ell$ the angular momentum number (we
decompose the field in spherical harmonics), $\mu$ the mass of the SF
(in units for which $G=c=\hbar=1$), and $r_*$ ($-\infty < r_* <
+\infty$) the radial ``tortoise'' coordinate, related to the
Schwarzschild radial coordinate $r$ ($r > 2M$) by $r_* = r +
2M\ln(r/2M-1)$. As discussed in detail in~\cite{Burt:2011pv} (see also
reference~\cite{Grain:2007gn}), for each $\ell\geq 0$ and $M\mu <
M\mu_{\rm crit}(\ell)$, $V$ describes a potential well which is
enclosed between a barrier close to the BH, and the asymptotic
positive value $\mu^2$ of the potential at infinity.

The stationary solutions of Eq.~(\ref{KG.Pot}) are characterized by a
time dependency of the form $e^{i\omega t}$ with a real frequency
$\omega$.  Such solutions behave as a combination of ingoing and
outgoing waves to the left of the potential barrier, close to the
horizon, and we require them to decay exponentially to zero at spatial
infinity.  There is a continuous spectrum of such solutions for
$\omega^2 < \mu^2$.  Furthermore, as we have shown
in~\cite{Burt:2011pv}, for $\omega$ lying in the region $V_{\rm min}
<\omega^2 < \mu^2$, with $V_{\rm min}$ the local minimum of the
potential, there is a discrete set of such frequencies for which the
amplitude inside the potential well takes very large values when
compared with the amplitude close to the horizon. We call these modes
the {\it stationary resonances}, and the region $V_{\rm min} <\omega^2
< \mu^2$ the {\em resonance band}.

If one imposes the condition of no waves coming from the region close
to the horizon, while keeping the requirement of exponential decay at
spatial infinity, it turns out that Eq.~(\ref{KG.Pot}) is only
satisfied for a discrete set of complex frequencies. These solutions
have been called {\em quasi-resonances} in the
literature~\cite{Ohashi:2004wr} (see also~\cite{Detweiler1980} for a
previous study of such solutions).

Both the stationary resonances and quasi-resonances are in fact
non-physical solutions. First, as mentioned above, all purely
stationary solutions, {\em i.e.} those with real $\omega$, require
waves to move outward from the horizon region to compensate for the
waves that tunnel out through the barrier and move toward the horizon
(as otherwise the situation would not be stationary).  Imposing the
condition of no waves coming out from the horizon clearly improves the
situation at the cost of introducing complex valued frequencies
$\omega$.  This makes sense physically since now the solutions must
decay in time as the waves tunnel out of the potential well and fall
toward the BH, with this decay represented by the imaginary part of
the frequency.  Nevertheless, it turns out that such solutions are
still non-physical, as one can show that the energy density diverges
at the horizon for both types of solutions.

There is yet a third class of resonant solutions we will consider.  In
reference~\cite{Burt:2011pv} we performed numerical evolutions of
regular initial data that has not such divergence of the energy
density at the horizon, and found damped oscillating solutions that
remain surrounding the BH for long times. These solutions have
frequency equal (within the numerical error) to that of the stationary
resonances.  We call these solutions {\em dynamical resonances}.  We
will show that the real part of the frequency of the quasi-resonant
modes coincides with the frequency of oscillation of the stationary
and dynamical resonances, and that the imaginary part coincides with
the decay rate of the dynamical resonances. This is a nontrivial
result since, after all, the dynamical solutions are in some sense
``infinitely different'' to both the stationary resonances and the
quasi-resonances because they are regular solutions with finite
energy.  This is similar to the case of the quasi-normal modes of BHs,
which also diverge at the horizon, but nevertheless physical
excitations behave as combinations of them locally at late times.


{\it III- Toy model.}  In order to illustrate how the different
resonant modes arise and are related to each other, we first consider
a simple toy model in which the potential $V$ in
Eq.~(\ref{Eq:KGPotential}) is replaced by
\begin{equation}
U(x) = A\delta(x) + \mu^2\Theta(x-a) \;, \qquad x:=r_* \;, \label{potentialU}
\end{equation}
with $A$ and $a$ positive constants. That is, we have a potential well
enclosed between the $\delta$-barrier and a step function where $U$
jumps from zero to its asymptotic value $\mu^2$. For $M\mu\ll 1$ and
$\ell\geq 1$ the minimum of the potential well of the physical
potential $V$ is located at \mbox{$r_{\rm min} =
  \ell(\ell+1)(M\mu^2)^{-1}[1 + {\cal O}(M\mu)^2]$}, and the strength
of its barrier is $\int_{-\infty}^{r_{\rm *min}} V dr_* =
[6\ell(\ell+1) + 1](4M)^{-1}[1 + {\cal O}(M\mu)^2]$. Therefore, we
require the dimensionless parameters $\lambda := 2\mu/A$ and $p :=
(a\mu)^{-1}$ to scale like $M\mu$ in our toy model potential $U$. In
the analysis below we prove that for fixed $p < 1/\pi$ and small
enough values of $\lambda$, the potential $U$ gives rise to a discrete
set of stationary and quasi-resonant frequencies whose real parts
agree with each other.

The resonant modes for the potential $U$ have the form
\begin{equation}
\phi(t,x) = e^{i\omega t}\cdot\left\{ \begin{array}{rl}
\alpha e^{i\omega x} + \beta e^{-i\omega x}, & \; x < 0\; ,\\
\gamma e^{i\omega x} + \delta e^{-i\omega x}, & \; 0 < x < a\; ,\\
e^{-\sqrt{\mu^2 - \omega^2} x}, & \; x > a\; ,
\end{array} \right.
\end{equation}
with $\omega$ a complex frequency. We require $\omega\notin
(-\infty,-\mu]\cup [\mu,\infty)$, which guarantees that
$\re\sqrt{\mu^2-\omega^2}\neq 0$, and therefore we can choose the
sign of the square root such that its real part is positive,
leading to exponential decay for $x > a$, as required. The
constants $\alpha$, $\beta$, $\gamma$ and $\delta$ are determined by
the matching conditions at $x=0$ and $x=a$ which consist in
the continuity of $\phi$ and its first spatial derivative $\phi'$ at
$x=a$ and the continuity of $\phi$ and the jump condition
$\phi'(t,0^+) - \phi'(t,0^-) = A\phi(t,0)$ at $x=0$.

For the stationary resonances $\omega$ is real and our restrictions
imply $\omega\in (-\mu,\mu)$. Since $U$ is real, it follows in this
case that $\alpha = \beta^*$, $\gamma = \delta^*$, and the ratio
between the amplitudes of the solution for $x < 0$ and the one inside
the well is $\nu(\omega) = |\alpha|/|\gamma|$. The resonance
frequencies are the ones that minimize this ratio. For the
quasi-resonances, in turn, the condition of no waves coming from the
horizon implies that $\beta=0$, which can only occur for a nontrivial
imaginary part of $\omega$. In terms of the quantities $\lambda$ and
$p$ defined below Eq. (\ref{potentialU}), and the complex angle
$\varphi$ lying inside the strip $-\pi/2 < \re(\varphi) < \pi/2$
defined by $\omega = \mu\sin\varphi$, the condition $\beta=0$ reads
\begin{equation}
F_p(\lambda,\varphi) 
 := 1 + i\lambda\sin\varphi - e^{-2i\varphi - \frac{2i}{p}\sin\varphi} = 0\, ,
\label{Eq:Zeroes}
\end{equation}
while the ratio of interior and exterior solutions in the stationary
case is
\begin{equation}
\nu(\omega) = \frac{1}{\lambda}\frac{|F_p(\lambda,\varphi)|}{\sin\varphi} \; .
\label{Eq:Ratio}
\end{equation}
As we prove now there exists, for each $p < 1/\pi$ and small enough
$\lambda$, a finite family $\varphi_n(\lambda)$ of solutions of
Eq.~(\ref{Eq:Zeroes}) with the following properties: they depend
smoothly on $\lambda$, their imaginary parts decay like $\lambda^2$,
and their real parts agree with the local minima of $\nu(\omega)$.
This shows the existence of quasi-resonant modes for the toy model
potential $U$ whose frequencies are related to those of the stationary
resonant modes through their real parts.

In order to show this we consider first the limit \mbox{$\lambda=0$},
in which case Eq.~(\ref{Eq:Zeroes}) yields real solutions
$\varphi_n^{(0)}\in (-\pi/2,\pi/2)$ satisfying $\sin\varphi_n^{(0)} =
p(n\pi - \varphi_n^{(0)})$, with $n=1,2,3,\ldots$. There are a finite
number of solutions when $p < 1/\pi$. For $p\ll 1/\pi$ the fundamental
solution is approximately $\varphi^{(0)}_1\approx p\pi$. In order to
extend these solutions for small $\lambda > 0$ we invoke the implicit
function theorem and observe that the function $F_p(\lambda,\varphi)$
is linear in $\lambda$ and analytic in $\varphi$. Furthermore,
$\frac{\partial F_p}{\partial\varphi}(0,\varphi_n^{(0)}) = 2i(1 +
p^{-1}\cos\varphi_n^{(0)}) \neq 0$ since $-\pi/2 < \varphi_n^{(0)} <
\pi/2$. Therefore, there exists for small enough $\lambda$ a unique
smooth solution curve $\varphi_n(\lambda)$ satisfying
$F_p(\lambda,\varphi_n(\lambda)) = 0$ and $\varphi_n(0) =
\varphi_n^{(0)}$. A Taylor expansion reveals that the corresponding
complex frequencies $\omega_n(\lambda) = \mu\sin(\varphi_n(\lambda))$
have the form
\begin{subequations}
\begin{eqnarray}
\re(\omega_n(\lambda))
 &=& \mu\sin\varphi_n^{(0)}\left[ 1 - \frac{\lambda}{q_n} 
  + \frac{\sigma}{q_n^2}\lambda^2 + {\cal O}(\lambda^3) \right], \hspace{5mm}
\label{Eq:Reomegan}\\
\im(\omega_n(\lambda))
 &=& \frac{\mu}{2}\frac{\sin^2\varphi_n^{(0)}}{q_n}
 \lambda^2 + {\cal O}(\lambda^3) \; ,
\label{Eq:Imomegan}
\end{eqnarray}
\end{subequations}
where we have defined $q_n := 2(\cos^{-1}\varphi_n^{(0)} + p^{-1})$
and $\sigma := 1 -
q_n^{-1}\sin^2\varphi_n^{(0)}/\cos^3\varphi_n^{(0)}$.  Furthermore,
one can verify that the local minima of the function $\nu(\omega)$
given in Eq.~(\ref{Eq:Ratio}), describing the stationary resonances,
have precisely the expansion given in
Eq.~(\ref{Eq:Reomegan}). Therefore, up to second order in $\lambda$,
the frequencies of the stationary resonances agree with the real part
of the quasi-resonant frequencies.

Since both parameters $\lambda$ and $p$ scale like $M\mu$ in our model
potential, it follows that for $M\mu\ll 1$ the imaginary part of
$M\omega_n$ scales like $(M\mu)^6$.  This explains, at least
qualitatively, why the quasi-resonant modes are associated with such
large time-scales when $M\mu\ll 1$.

It is also possible to relate the quasi-resonant frequencies
$\omega_n$ to the dynamical resonances by representing the solutions
of Eq.~(\ref{KG.Pot}) as an inverse Laplace integral. This calculation
will be presented elsewhere.

\begin{figure}[ht]
\includegraphics[angle=0,width=0.49\textwidth,height=!,clip]{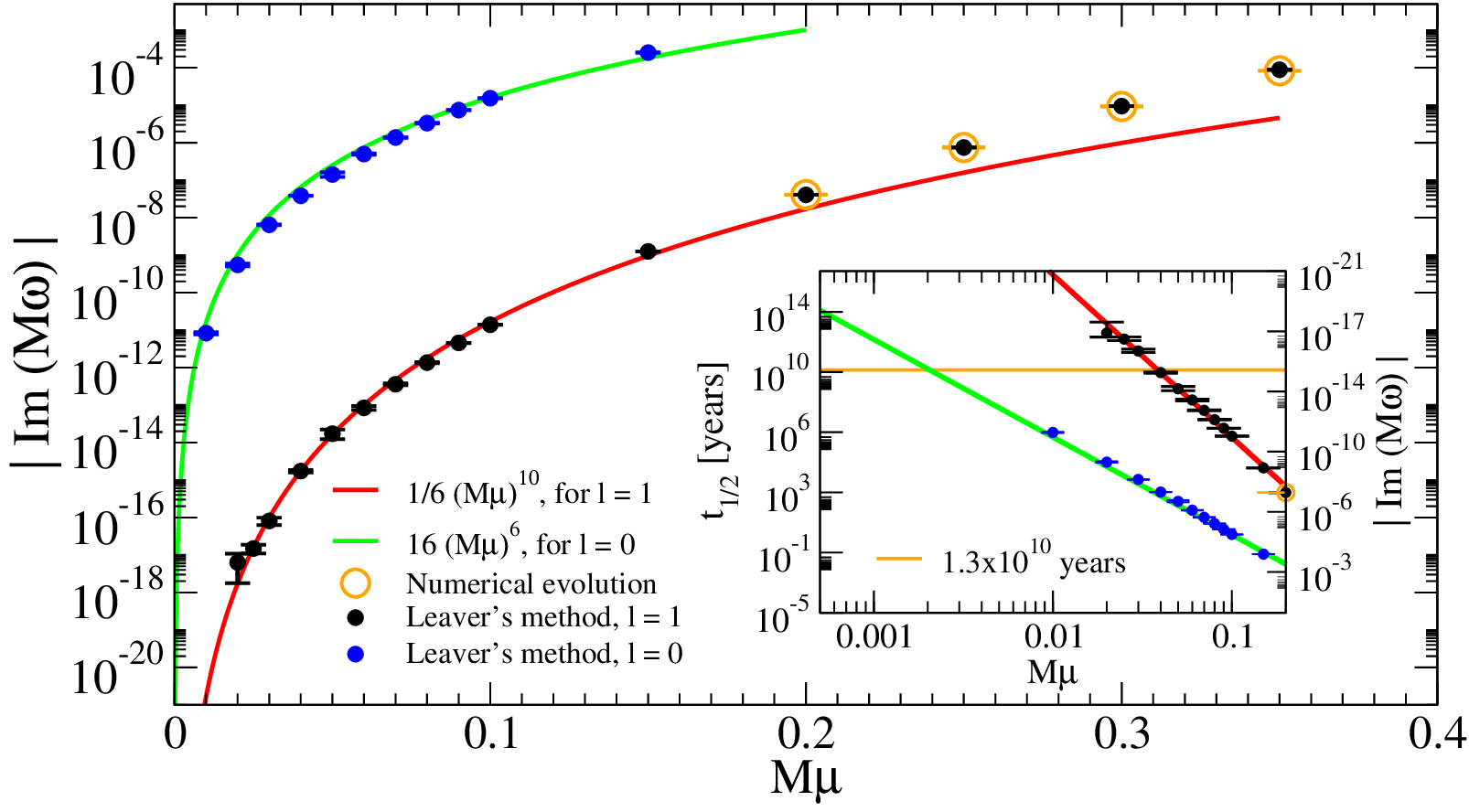}
\caption{The imaginary part of the frequency for the first
  quasi-resonant mode obtained using Leaver's method is shown as
  function of $M \mu$ for the cases $\ell=0,1$.}
\label{f:mu_vs_Imwlog}
\end{figure}

{\it IV- Frequency comparison and characteristic times.}  For the
physical potential, Eq. (\ref{Eq:KGPotential}), the quasi-resonant
modes can be obtained semi-analytically in several ways, the most
common are the continued fraction method introduced by Leaver
\cite{Leaver:1985ax}, and the WKB approach \cite{Iyer:1986np} (see
\cite{Nollert:1999ji, Kokkotas99a, Konoplya:2011qq, Berti:2009kk} for
details).  An analytic expression valid for small values of the
combination $M\mu$ was obtained by Detweiler in
\cite{Detweiler1980}. The stationary and dynamical modes were
previously considered in~\cite{Burt:2011pv}.  In this section we show
that the oscillations of the dynamical resonances have frequencies
equal (up to numerical error) to the real part of the frequencies of
the quasi-resonant modes, while the decay rates are equal to the
imaginary part of such frequencies (see Fig. \ref{f:mu_vs_Imwlog}). We
also show that the imaginary part of the quasi-resonant frequencies
calculated by means of Leaver's method has the limit behaviour
predicted by Detweiler. Hence, we can use Detweiler expressions to
determine the decay rate of the dynamical resonances with very small
values of $M\mu$, which cannot be reached with the other two methods.

Leaver's semi-analytical approach \cite{Leaver:1985ax} assumes a
harmonic time dependence for the SF, $\phi(t,r)=\psi(r)e^{i\omega t}$,
with $\omega$ the complex quasi-resonant frequency, and proposes a
power series expansion for the radial function $\psi(r)$ of the form
\begin{equation}
\psi = \tilde\psi \sum_{n=0}^{\infty} a_{n}\left( 1-\frac{2M}{r} 
\right)^n \;,
\end{equation}
where, following Ref.~\cite{Konoplya:2004wg}, we define $\tilde \psi
:= e^{-\chi r}r^{(-2\,M\chi + M\mu^2/\chi)}\left( 1-2M/r
\right)^{-2iM\omega}$.  This guarantees that close to the horizon
$\psi \sim e^{i\omega r^{*}}$, whereas at spatial infinity $\psi \sim
e^{-\chi r^{*}}r^{M\mu^{2}/\chi}$, with $\chi = \sqrt{\mu^2-\omega^2}$
and the sign convention for the square root was explained in the
previous section. Notice however that our boundary conditions are
different than those in \cite{Konoplya:2004wg}.

In Fig.~\ref{f:mu_vs_Imwlog} we plot the imaginary part of the
frequency of the first quasi-resonant mode (the one with the lower
value of $\textrm{Re}(M\omega)$) for the cases $\ell=0,1$ and
different values of $M\mu$.  We find that the real part of the
frequencies obtained corresponds to the frequencies at which the SF
oscillates in a dynamical scenario~\cite{Burt:2011pv}. Furthermore, we
also find that the imaginary part of the frequencies (dots in
Fig.~\ref{f:mu_vs_Imwlog}) corresponds to the decay rate of the field
in the numerical evolutions (empty circles in
Fig.~\ref{f:mu_vs_Imwlog}).

In the numerical evolutions we were not able to evolve the
configurations for small values of $M\mu$, both because of the
propagation of the errors in the numerical approximation, and because
the time required for such evolutions becomes prohibitive.  Even
though Leaver's method allows us to obtain results for parameter
values that could not be reached with the numerical evolutions,
numerical roundoff errors make it still prohibitive to obtain
accurately very small values for the imaginary part of the
quasi-resonant frequencies.

In reference~\cite{Detweiler1980} analytic expressions for the
spectrum of quasi-resonant modes were found in the limit $M\mu\ll 1$,
for values of $\ell \ge 1$.  For $\ell=1$, the imaginary part of the
frequency for the first quasi-resonant mode in this limit 
is given by
$\textrm{Im}(M\omega) = (M\mu)^{10}/6$.
Strictly speaking, the expressions given in~\cite{Detweiler1980} may
not apply for the case with $\ell=0$. However, if we nevertheless use
them in that case we find $\textrm{Im}(M\omega) = 16(M\mu)^{6}$.  We
have calculated the quasi-resonant mode frequencies using Leaver's
method up to the smallest values of $M\mu$ allowed by the roundoff
errors, and we have found that the imaginary part of such frequencies
matches the Detweiler approximations (within errors) at small values
of $M\mu$ (see Fig.~\ref{f:mu_vs_Imwlog}).  This result allows us to
conclude that, for small values of $M\mu$, the decay time of the
dynamical resonances should correspond to the one predicted by
Detweiler in~\cite{Detweiler1980}.

\begin{figure}[ht]
\includegraphics[angle=0,width=0.49\textwidth,height=!,clip]{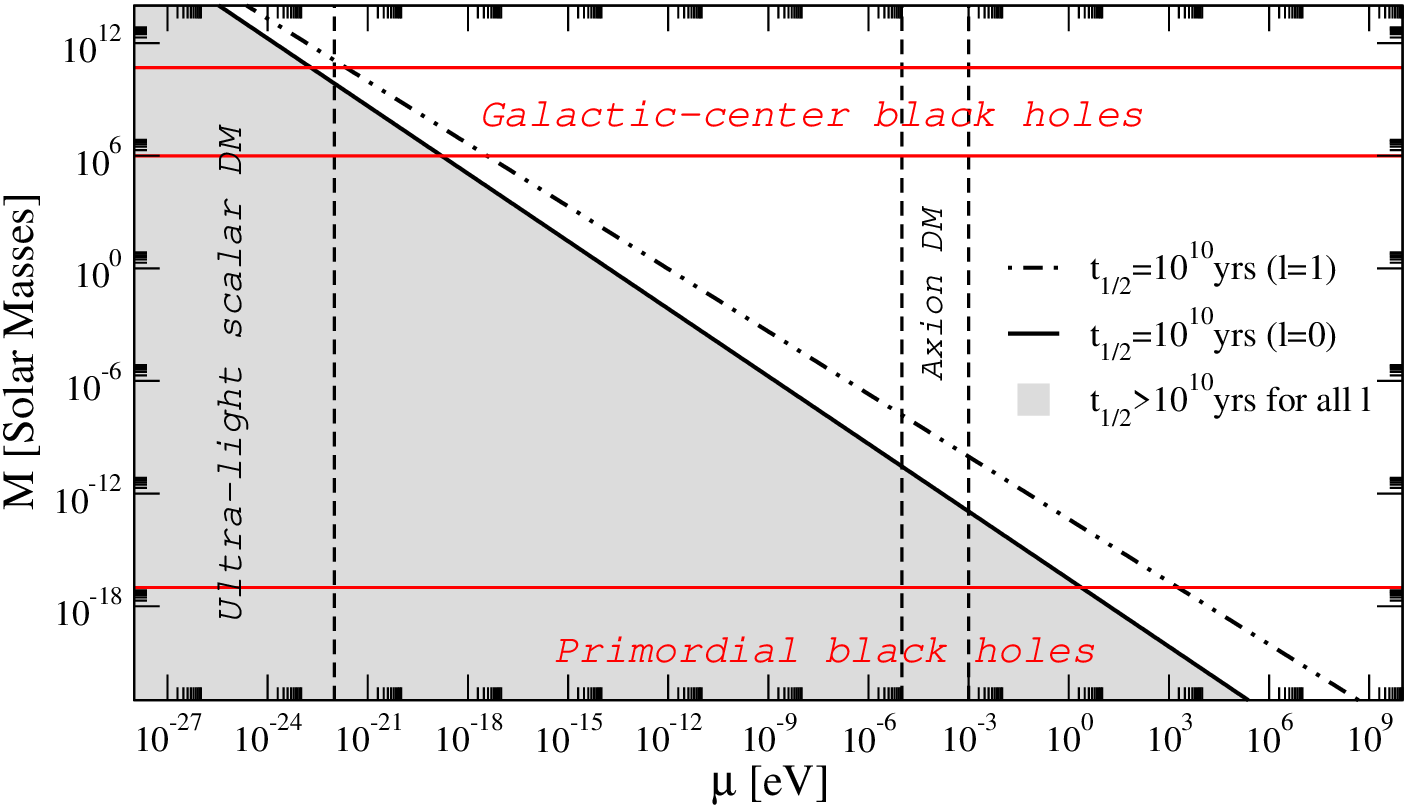}
\caption{Combinations of the parameters $M$ and $\mu$ for which the SF
  half-life time can be larger than the age of the universe.}
\label{f:mu_vs_M_survival}
\end{figure}

Since the decay rate of the dynamical resonances is related to the
imaginary part of the quasi-resonant frequencies, then its half-life
time is inversely proportional to $\textrm{Im} (M\omega)$. In
particular we can find BH and SF masses such that $t_{1/2} \sim
10^{10}$ yrs. and larger for the cases $\ell=0,1$ previously
mentioned.  Those results are shown in Fig.  \ref{f:mu_vs_M_survival}.
There are two distinct regions of the parameter space of physical
interest for which the configurations live longer than the age of the
Universe: (i) a SF mass smaller than $1$~eV and BH mass smaller than
$10^{-17}M_\odot$, consistent with primordial BHs with an axion
distribution~\cite{Sikivie:2009qn}; and (ii) an ultra-light (fuzzy)
SF~\cite{Hu:2000ke, Matos:2000ss, Boehmer:2007um} with mass smaller
than $10^{-22}$~eV and a supermassive BH with mass smaller than
$5\times10^{10}M_\odot$, as could be the case for a dark matter
halo~\cite{Sin:1992bg, Arbey:2001qi} surrounding a BH at a galactic
center.

{\it V- Evolution of arbitrary configurations.}  In this section we
study the long-term numerical evolution of SF distributions that are
initially surrounding a Schwarzschild BH.  We show that at late times,
after some SF falls into the BH and some escapes to infinity, the
remaining SF consists of a superposition of the dynamical resonances,
which, for practical purposes, can be described in terms of the
quasi-resonant modes.  Remarkably, we see that this holds even for
pretty arbitrary initial data.  Note that the calculations in this
section are identical to those in~\cite{Burt:2011pv}, but with the
crucial difference that here they are performed on arbitrary initial
data (as opposed to initial data obtained from stationary resonances),
hence leading to new, much stronger results.

In order to test the evolution of a variety of SF
distributions, we construct a two-parameter family of initial data, of
the form
\begin{equation}
u_0(r) = \begin{cases}
  N (r-R_1)^4(r-R_2)^4 & \text{for $R_1 \leq r \leq R_2$} \\
  0 & \text{otherwise}
\end{cases}\;, \label{eq:id}
\end{equation}
with the normalization $N=[2/(R_1-R_2)]^8$.  The parameters $R_1$ and
$R_2$ are chosen in order to set distributions at $t=0$ with different
``locations'' and ``sizes''.  The initial value of the time
derivative, $\left.\dot{u}\right|_{t=0}$, is constructed as
\begin{equation}
\left.\dot{u}\right|_{t=0} = \left.\frac{\partial f}{\partial t}\right|_{t=0}\;, \quad
f(t,r) := u_0(r-vt)\;,
\end{equation}
where $v$ is a free parameter that has no significant effect on the
results presented here, and for the majority of configurations studied
is simply set to zero.  The evolution equation is solved numerically,
as described in~\cite{Megevand:2007uy}.

As expected, during the initial stages of the evolution some SF
accretes into the BH, while some is radiated away, both with rates
that depend greatly on the initial data chosen, and can be very large
in some cases. However, at late times, all evolutions show a similar
steady behaviour with slow accretion into the BH.  Similar results are
obtained when studying the long-term evolution of a variety of
configurations, some of them very different in size and spatial
distribution $-$although all distributions studied here are quite
``wide'' when compared to the BH size.

\begin{figure}[ht]
\includegraphics[angle=0,width=0.48\textwidth,height=!,trim=0 1mm 0 0,clip=true]{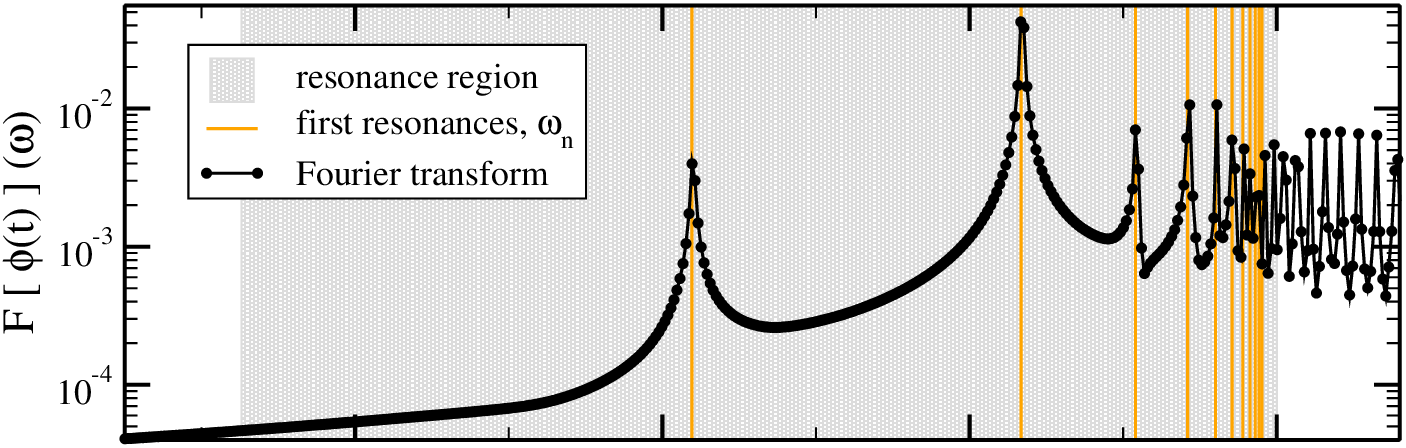}
\includegraphics[angle=0,width=0.48\textwidth,height=!,trim=0 0 0 0.6mm,clip=true]{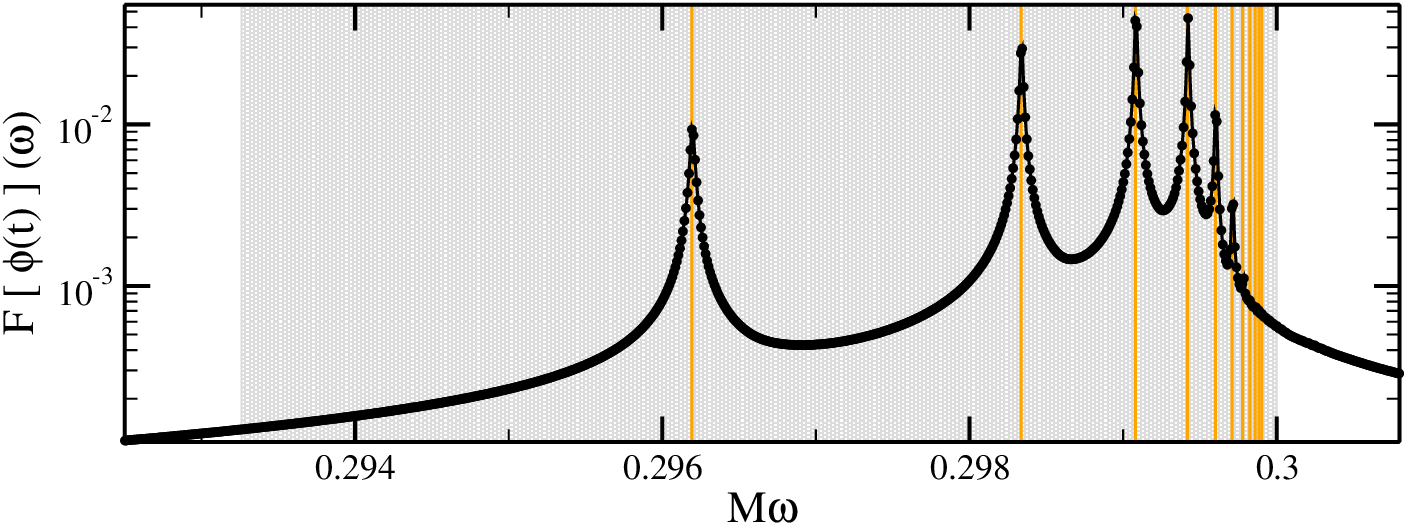}
\caption{Fourier transform in time of the evolution of initial data
  with $R_1=100$, $R_2=200$ (top panel); and $R_1=-1000$, $R_2=1000$
  (bottom panel).  The peaks to the right of the resonance band in the
  top panel are a numerical artifact caused by noise originating at
  the outer boundary.}
\label{f:fourier2}
\end{figure}

One can gain some understanding of this behaviour by conducting a
spectral analysis as follows.  We perform a (discrete) Fourier
transform in time, $\mathcal{F}[\phi(t,r_j)](\omega),$ of the SF at
several (about 20) sample points $r_j$, then take the average to
obtain the spectrum $\mathcal{F}[\phi(t)](\omega)$.  For brevity, we
present here results corresponding to only two representative
examples.  They correspond to initial data with $R_1=100M$,
$R_2=200M$; and $R_1=-1000M$, $R_2=1000M$.  In both cases $\mu M=0.3$,
$\ell=1$ and $v=0$.  The spectra are shown in Fig.~\ref{f:fourier2}.

For comparison, we have indicated in the figure the first resonance
frequencies (vertical lines), as obtained in~\cite{Burt:2011pv}.  One
can clearly see peaks that coincide with each resonant frequency.
These results seem to indicate that, at late times, even quite generic
SF distributions evolve as a combination of the resonant modes, which,
as we have shown, can last for cosmological time-scales.


This work was supported in part by CONACyT grants 82787 and 167335,
DGAPA-UNAM through grant IN115311, and SNI-M\'exico. JCD and MM
acknowledge DGAPA-UNAM for postdoctoral grants.  OS was also supported
by grant CIC 4.19 from Universidad Michoacana.  This work is part of
the ``Instituto Avanzado de Cosmolog\'ia'' collaboration.



\end{document}